\documentclass[%
 aip,
 amsmath,amssymb,
 reprint,%
]{revtex4-1}

\newcommand{\uvec}[1]{\boldsymbol{\hat{\textbf{#1}}}}

\usepackage{siunitx}
\usepackage{graphicx}
\usepackage{dcolumn}
\usepackage{bm}
\usepackage{dsfont}
\usepackage{color}
\usepackage[utf8]{inputenc}
\usepackage[T1]{fontenc}
\usepackage{mathptmx}
\usepackage{float}

\begin{document}

\preprint{AIP/123-QED}

\title[Chaplain and Craster]{Surface corrugated laminates as elastic grating couplers: splitting of SV- and P-waves by selective diffraction}

\author{G. J. Chaplain}
\affiliation{Department of Mathematics, Imperial College London, London SW7 2AZ, UK}
\email{gregory.chaplain16@imperial.ac.uk}
\author{R. V. Craster}%
\affiliation{Department of Mathematics, Imperial College London, London SW7 2AZ, UK}%
\affiliation{Department of Mechanical Engineering, Imperial College London, London SW7 2AZ, UK }
\affiliation{UMI 2004 Abraham de Moivre-CNRS, Imperial College London, London SW7 2AZ, UK}


\date{\today}

\begin{abstract}
The phenomenon of selective diffraction is extended to in-plane elastic waves and we design surface corrugated periodic laminates that incorporate crystal momentum transfer which, due to the rich physics embedded within the vector elastic system, results in frequency, angle and wave-type selective diffraction. The resulting devices are elastic grating couplers, with additional capabilities as compared to analogous scalar electromagnetic couplers, in that the elastic couplers possess the ability to split and independently redirect, through selective negative refraction, the two body waves present in the vector elastic system: compressional (P) and shear-vertical (SV) elastic waves. The design paradigm, and interpretation, is aided by obtaining isofrequency contours via a non-dimensionalised transfer matrix method. 
\end{abstract}

\maketitle

\section{\label{sec:intro} Introduction}
Numerous parallels exist between the in-plane vector elastic wave system and the scalar wave systems of polarised electromagnetism, acoustics, flexural waves, linear water waves and shear-horizontal polarised elasticity. Transposing physics into the vectorial elastic system is often nuanced and not straightforward with different shear and compressional wavespeeds in elasticity, and coupling between shear and compression at interfaces, that can lead to additional effects. An example being array-guided surface waves confined to elastic diffraction gratings, so-called Rayleigh-Bloch waves \cite{colquitt2015rayleigh}, that are elastic analogues to electromagnetic spoof surface plasmon polaritons\cite{pendry04a,hibbins2005experimental}, which solely owe their existence to periodic structuring. The nature of the vector elastic system dictates that these modes exist for traction-free boundary conditions, rather than simple Neumann conditions as in, for example, electromagnetic gratings \cite{wilcox1984scattering}. In two-dimensions the nuances are driven in the vector elastic system by the presence of two body waves, shear-vertical polarised (SV) and compressional (P), that co-exist and travel with different wavespeeds $c_s$ and $c_p$ respectively, such that $c_s < c_p$. Additionally these wave types couple to each other (and can scatter into  surface Rayleigh waves if defects are present) at traction-free interfaces\cite{graff75a}, and therefore they are inherently coupled and resist independent manipulation. Since the advent of metamaterials there has been considerable effort to manipulate these distinct elastic wave types, drawing on techniques developed in optics and  electromagnetism, particularly those associated with the abrupt phase changes endowed to wavefronts which traverse a particular metasurface\cite{yu2014flat}. Anomalous reflection, refraction, transmission and wave splitting effects have been proposed, by designing metamaterials and metasurfaces based on phased gratings; spatially dependent structurings give the required tailored phase profile, allowing generalised Snell's laws to be observed\cite{su2016focusing,su2018elastic,ahn2019topology}.

\begin{figure}[H]
    \centering
    \includegraphics[width = 0.35\textwidth]{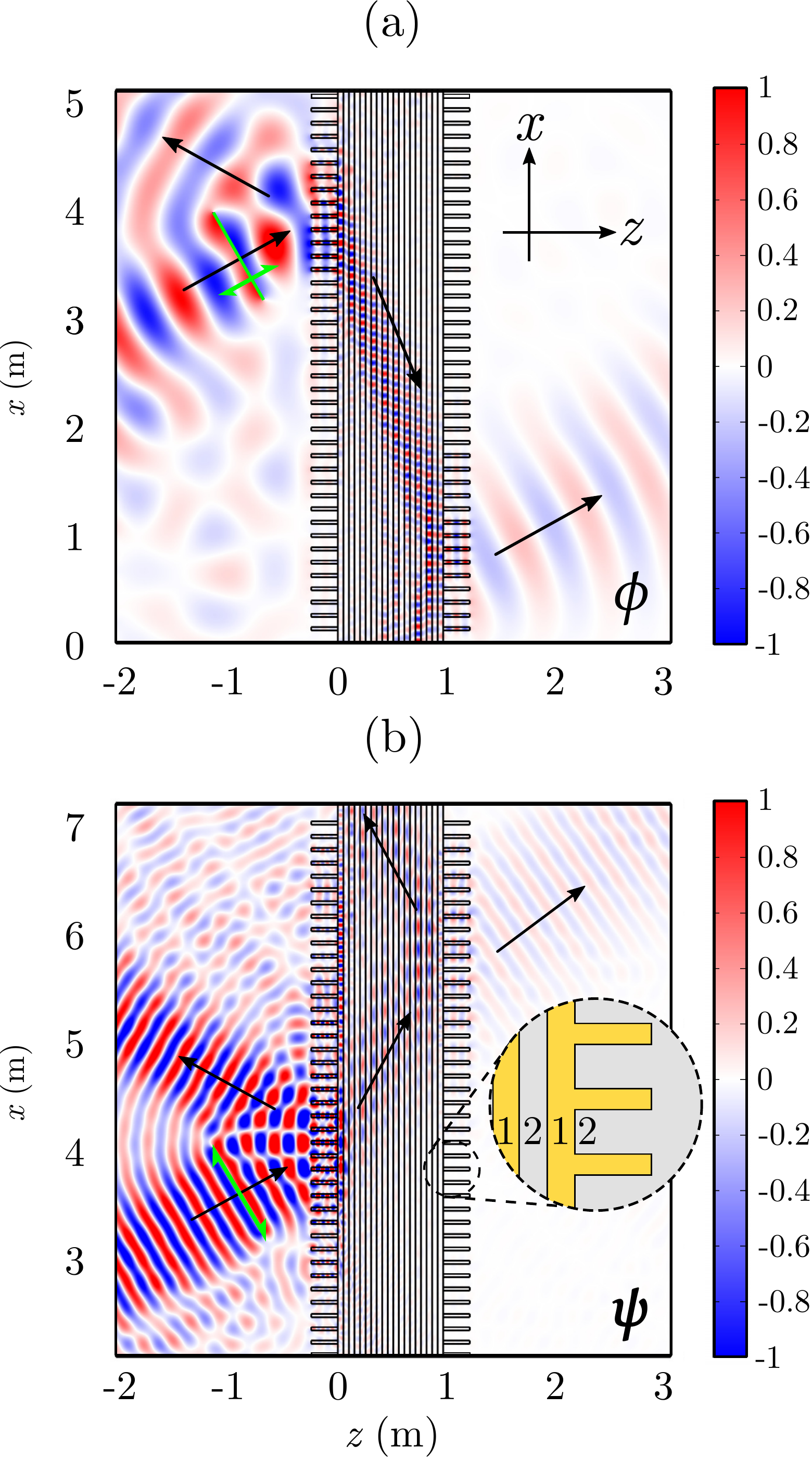}
    \caption{Splitting of SV- and P- elastic waves by the selective diffraction created from surface corrugation decorating an elastic bilayer laminate: (a) Compressional elastic potential field, $\phi$. (b) Shear-vertical potential field, $\boldsymbol{\psi}$. Each are excited by separate line forcings (green line) which match the relative preferential particle motion direction of each wave type (half-headed arrows). The ray path is shown by the black arrows. The two wave types are spatially separated upon transmission through the laminate; only the compressional P-waves receive a momentum "kick" due to the surface corrugation, resulting in selective negative refraction. Geometry detailed in Appendix~\ref{sec:Append1}.}
    \label{fig:Hero}
\end{figure}

We extend conventional grating coupler concepts
from electromagnetism and optics using periodic decorations attached to an elastic laminate, see Fig. \ref{fig:Hero}, by leveraging momentum transfer to generate a selective momentum "kick". In optics the generation of a momentum "kick" has been used to create negative refraction \cite{Lu2007}, however due to the scalar nature of polarised electromagnetism there is no  possibility for the analogues of compressional and shear wave splitting. 

By  using isofrequency contours as a design tool these surface corrugations on the outer interfaces of an elastic laminate, composed of periodic bilayers analogous to one-dimensional dielectric photonic crystals\cite{joannopoulos08a}, we are able to give the "kick" preferentially to one wave type and create selective negative refraction (in Fig. \ref{fig:Hero} the kick is gifted to the compressional wave); the presence of the two distinct elastic body wave types allows these extended coupler concepts to provide flexible control over P and SV waves independently. Figure~\ref{fig:Hero} motivates these additional capabilities of the elastic system showing the compressional and shear elastic potentials $\phi$ and $\boldsymbol{\psi}$; the finite element method is used in COMSOL Multiphysics to perform the simulations. To emphasise the different effects we perform multiple scattering simulations, with alternate line forcings, to preferentially excite either P or SV-waves; dominant compressional motion is achieved when the forcing direction is perpendicular to the line source, whilst shear motion is dominant when the forcing direction is parallel to it. This enhances visualisation of the effects by distinguishing between the mixture of the two wave types which result from a general source (Appendix~\ref{sec:AppendB} highlights an example where the total solid displacement field is visualised). The forcing directions are highlighted by the green half-headed arrows in Figure~\ref{fig:Hero}, which follow the particle motion. 
 

Alongside the additional physics of wave splitting for elasticity, more conventional grating coupler effects, analogous to those from electromagnetism, are also recoverable. Indeed the term grating coupler is almost exclusively used to describe electromagnetic devices, particularly prevalent in integrated silicon photonic devices and fibre optics \cite{taillaert2002out,chuang2012physics,marchetti2017high,marchetti2019coupling}. The desired outcome in these systems is to couple incident light efficiently into a waveguide, and ensure its confinement. We show, as an additional example in Appendix~\ref{sec:AppendB}, that this is entirely possible for elastic laminate waveguides by suitable design of the surface corrugation.


The structure design rests on obtaining isofrequency contours for the elastic laminate corresponding to purely P or SV-waves propagating through the laminate. We obtain those through employing the Transfer Matrix Method (TMM), as outlined in Section~\ref{sec:bilyaer}. Advantages of this methodology arise through its simplicity, permitting the design of structures with a tailored response over a wide range of frequencies and angles of incidence. Extending the concepts of periodic surface corrugation to elastic grating couplers thus allows the decoupling of elastic waves at an interface, by a simple alteration of the surface structure. The resulting control gained over each independent wave-type has practical implications in non-destructive evaluation and vibration isolation\cite{Chimenti}. 

\section{\label{sec:bilyaer} Periodic Bilayer Laminates and the Transfer Matrix Method}

Elastic waves in isotropic, homogeneous layers are governed by the elastic wave equation\cite{graff75a}
\begin{equation}
 \rho\boldsymbol{\ddot{u}} = (\lambda+2\mu)\nabla(\nabla\cdotp\boldsymbol{u}) -  \mu\nabla\times\nabla\times\boldsymbol{u},
 \label{eq:linearelastic}
\end{equation}
where, in Cartesian coordinates, $\boldsymbol{u} = (u,v,w)$ is the displacement vector. The acceleration is given by the double time derivative, $\boldsymbol{\ddot{u}}$, with $\rho$, $\lambda$ and $\mu$ being the material density and Lam\'{e}'s first and second parameters respectively. Through Helmholtz decomposition, the displacement vector $\boldsymbol{u}$ can be written in terms of the divergence-less shear potential, $\boldsymbol{\psi}$, and the curl-less compressional potential, $\phi$, such that
\begin{equation}
    \boldsymbol{u} = \nabla\times\boldsymbol\psi + \nabla\phi.
\end{equation}
Time harmonic behaviour, of form $\exp({-i\omega t})$, is henceforth assumed. Following this, equation \eqref{eq:linearelastic} can be formulated such that each potential satisfies the wave equations
\begin{align}
(\nabla^2 +k_{p}^2)\phi = 0,\qquad
(\nabla^2 +k_s^2)\boldsymbol{\psi} = 0,
\end{align}
with $k_{p}$ and $k_{s}$ the compressional and shear wavenumbers related to the respective wavespeeds, $c_p, c_s$, through $k_{\delta} = \omega/c_{\delta}$, with $\delta = p, s$, such that $c_{p} = \sqrt{\frac{\lambda + 2\mu}{\rho}}$ and $c_{s} = \sqrt{\frac{\mu}{\rho}}$. The elastic potentials can then be written in terms of the displacement field through
\begin{align}
\phi = -\frac{(\nabla\cdotp\boldsymbol{u})}{k^2_p} = -\frac{\text{Tr}(\boldsymbol{\varepsilon})}{k^2_p}, \qquad
\boldsymbol{\psi} = \frac{(\nabla\times\boldsymbol{u})}{k^2_s}, 
\end{align}
where $\varepsilon$ is the strain tensor. The potentials provide a useful approach for separately visualising the compressional or shear components of an elastic wavefield.  

\begin{figure}
    \centering
    \includegraphics[width = 0.4\textwidth]{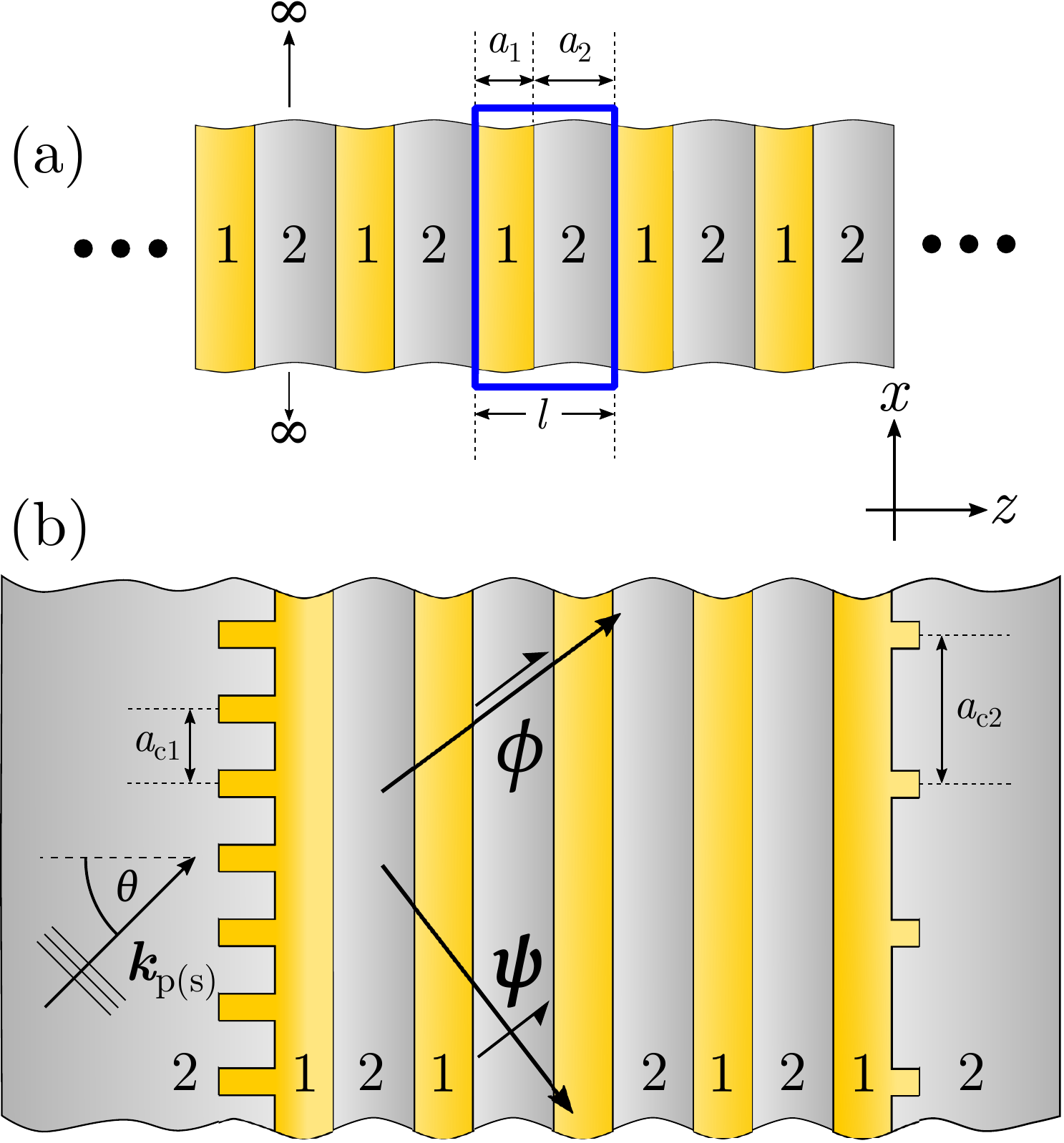}
    \caption{(a) Infinitely periodic bilayer elastic laminate: Periodic unit strips, of length $l = a_1+a_2$ where $a_1$ and $a_2$ are the thicknesses of elastic media labelled 1 and 2 respectively, comprise the laminate. The TMM is applied, along with Bloch conditions, across one unit cell, highlighted in blue. (b) Finite corrugated periodic bilayer elastic laminate: Surface corrugation takes place on the outermost layers of a finite laminate composed of the bilayers in (a). The vertical periodicity of the corrugations are defined by $a_{c1}$, $a_{c2}$ on the left and right sides of the laminate respectively, in the direction orthogonal to the laminate axis. In general $a_{c1} \neq a_{c2}$. Provided these periodicities are designed appropriately, for a given incident plane wave (P or SV) at angle $\theta$ to the laminate axis with wavevector $\boldsymbol{k}_{p(s)}$, the SV and P potential fields ($\boldsymbol{\psi}$ and $\phi$) can be separated by incorporating the momentum transfer by the grating. Half-headed arrows show particle motion. In each case the laminate is assumed to be infinite in the direction orthogonal to the periodicity. }
    \label{fig:Motivation}
\end{figure}

We first focus on a one-dimensional periodic bilayer elastic laminate structure, as shown in Figure~\ref{fig:Motivation}(a), that is of infinite extent and take advantage of periodicity to extract Floquet-Bloch dispersion curves that relate the Bloch wavenumber to frequency. The periodicity means that we can take a unit cell, and apply Bloch conditions; the unit strip is of width $l$, comprising two layers of linear elastic media, such that $l = a_1+a_2$, where $a_1, a_2$ are the widths of the first and second layers respectively. Following conventions in optics we define the direction of periodicity to be along the $z$-direction, with the strips assumed to be of infinite extent in the orthogonal $x$-direction. Each material has density $\rho$ along with Lam\'{e}'s first and second parameters $\lambda$, $\mu$ and  when required, the elastic quantities and parameters in each layer are distinguished with a subscript $i = 1,2$ for each layer.

To design finite gratings capable of performing selective diffraction, the isofrequency contours for a periodic array of the bilayer cell must be evaluated. Given the assumed time harmonic behaviour, the potential fields in each layer are written as a superposition of left- and right-propagating plane waves. To determine the structure's dispersive properties and band structure we employ the TMM, that relates the potential field amplitudes on either side of the unit cell and this is achieved by splitting the problem into two parts. The first requires the continuous physical quantities at an interface to be determined. In the elastic case considered here these are the normal and tangential displacements to the interface $(w,u)$, and the corresponding stresses $(\sigma_{zz},\sigma_{xz})$. As detailed in Appendix~\ref{sec:AppendA}, the vector of continuous quantities $\boldsymbol{W}_{i} = (w,u,\sigma_{zz},\sigma_{xz})^{\text{T}}$ across the interface within the $i^{th}$ layer is constructed through
\begin{align}
 \boldsymbol{W}_{i} = §
    \begin{pmatrix}
    \phi_{,z} - \boldsymbol{\psi}_{,x} \\ \\
    \phi_{,x} + \boldsymbol{\psi}_{,z} \\ \\
    (\lambda_{i} + 2\mu_{i})\nabla^2\phi + 2\mu_{i}(\phi_{,z} - \boldsymbol{\psi}_{,xz} - \nabla^2\phi) \\ \\
    \mu_{i}(2\phi_{,xz} + \boldsymbol{\psi}_{,zz} - \boldsymbol{\psi}_{,xx})
    \end{pmatrix},
\end{align}
where it is implicitly assumed that the potential fields correspond to those in the $i^{th}$ layer.

The second step relates the potential field amplitudes due to propagation through a layer, until the next interface is encountered, and therefore incorporating the phase change across the finite layer. By combining the two steps (using straightforward matrix multiplication) the amplitudes on one side of the unit strip are related to those on the other, through a matrix equation. 

Due to its relative simplicity, TMMs are often utilised to obtain transmission and reflection coefficients from finite structures, periodic or otherwise, (in any number of dimensions); for $N$ identical layers the resultant transfer matrix is simply raised to the $N^{th}$ power. As such this method has a long history, with much success in its applications to layered structures in electromagnetism \cite{markos2008wave}, solid-fluid interactions \cite{schoenberg1984wave}, disordered systems \cite{pernas2015scattering} and notably elasticity\cite{thomson1950transmission,haskell1953dispersion,huang1996oblique,fomenko2014plane}. By additionally invoking the quasi-periodic Floquet-Bloch conditions, the resultant matrix equation can be re-written as an eigenvalue problem which allows the dispersion relation to be obtained, with the Bloch wavenumber the eigenvalue for a given (non-dimensional) frequency $\Omega$, incidence angle and wave type. In Appendix~\ref{sec:AppendA} we detail this methodology, adopting a non-dimensional formalism; the matrices are composed of quantities expressed entirely in terms of the ratios of parameters of the two materials comprising the bilayer e.g. the ratios of the speeds in each material. The advantages of this are that we retain generality whilst mitigating numerical instabilities which arise through the required matrix inversion; the TMM is susceptible to instabilities in the so-called large \textit{f-d} problem\cite{dunkin1965computation}. Alternatives to the TMM are those based on the global matrix formulation\cite{knopoff1964matrix} or spectral collocation\cite{adamou2004spectral}.

\begin{figure}[h]
    \centering
    \includegraphics[width = 0.45\textwidth]{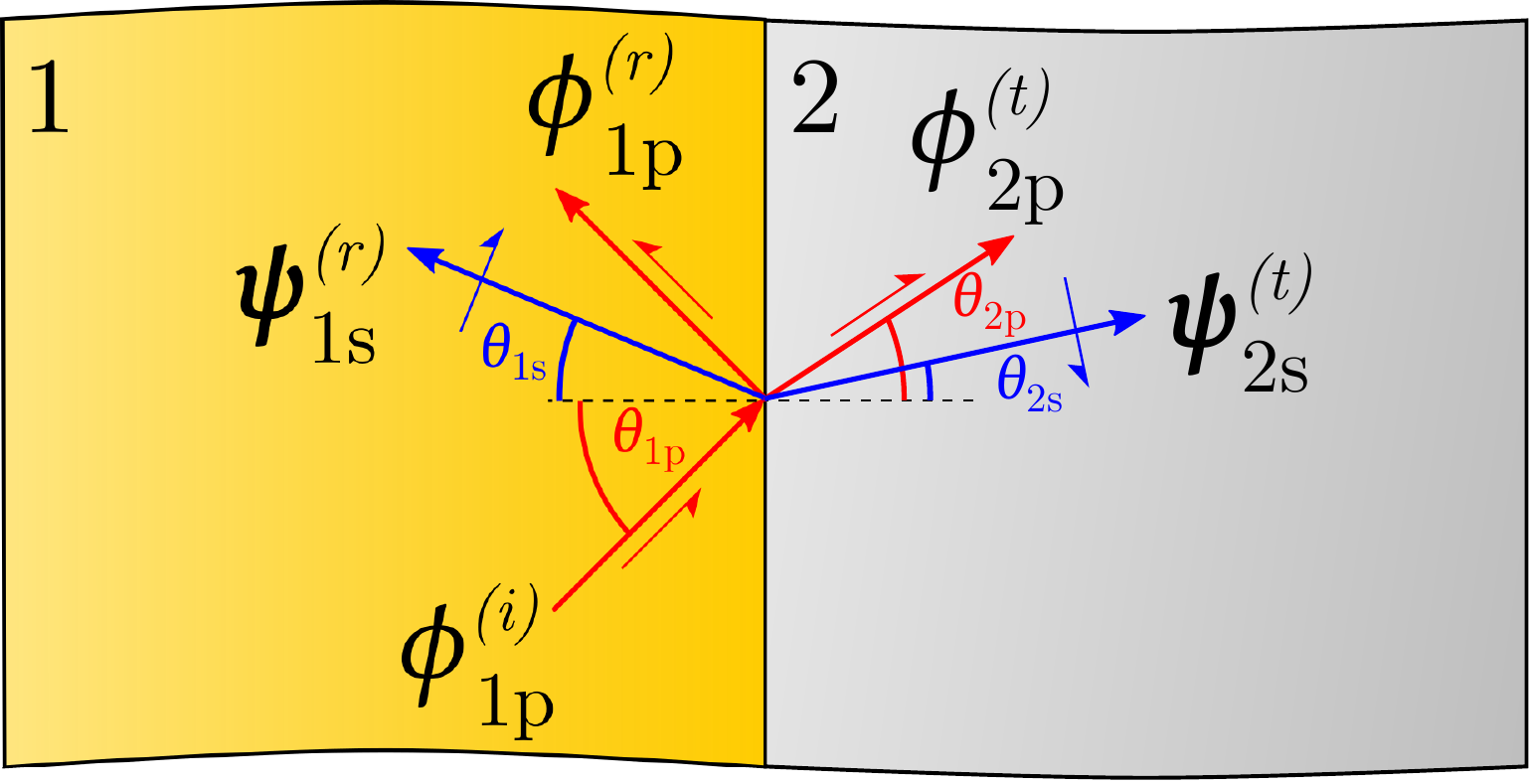}
    \caption{Snell's law for in-plane elasticity: An incident compressional wave in layer 1, described by the potential $\phi^{(i)}_{1p}$, couples to reflected and transmitted compressional waves $\phi^{(r)}_{1p}$ and $\phi^{(t)}_{2p}$ in layers 1 and 2 with wavespeeds $c_{1p}$ and $c_{2p}$ respectively. Additionally the incident compressional wave couples to reflected and transmitted shear-vertical potential fields $\boldsymbol{\psi}^{(r)}_{1s}$ and $\boldsymbol{\psi}^{(t)}_{2s}$ with wavespeeds $c_{1s}$ and $c_{2s}$ respectively. The angles of reflection and transmission are related to the incident angle through \eqref{eq:Snell}.}
    \label{fig:InterfaceCoupling}
\end{figure}

The method rapidly generates dispersion curves, and subsequently the isofrequency contours associated with the compressional or shear components of the incident wave. 
 Two independent calculations are required depending on the component of the incident wave under consideration; the angles of reflection and refraction are expressed in terms of the incident angle of either P- or SV-waves through Snell's law for in-plane elasticity\cite{achenbach2012wave}
\begin{equation}
    \frac{\sin\theta_{1p}}{c_{1p}} = \frac{\sin\theta_{1s}}{c_{1s}} = \frac{\sin\theta_{2p}}{c_{2p}} = \frac{\sin\theta_{2s}}{c_{2s}},
    \label{eq:Snell}
\end{equation}
where $\theta_{i\delta}$ and $c_{i\delta}$ give the angle and wavespeed corresponding to the P- and SV ($\delta = p,s$) waves in layers $i = 1,2$. A schematic of this coupling for an incident compressional wave, i.e. through $\phi$, is shown Figure~\ref{fig:InterfaceCoupling}.

Once the isofrequency contours of the infinitely periodic structure are obtained, these can be used to infer the behaviour of a finite laminate, given a large enough number of bilayers are taken\cite{joannopoulos08a}. The design of the corrugation for a particular effect then rests on interpreting these contours, as we do below.

\section{\label{sec:grating} Selective Diffraction through Surface Corrugation}
\begin{figure*}[t!]
    \centering
    \includegraphics[width = 0.85\textwidth]{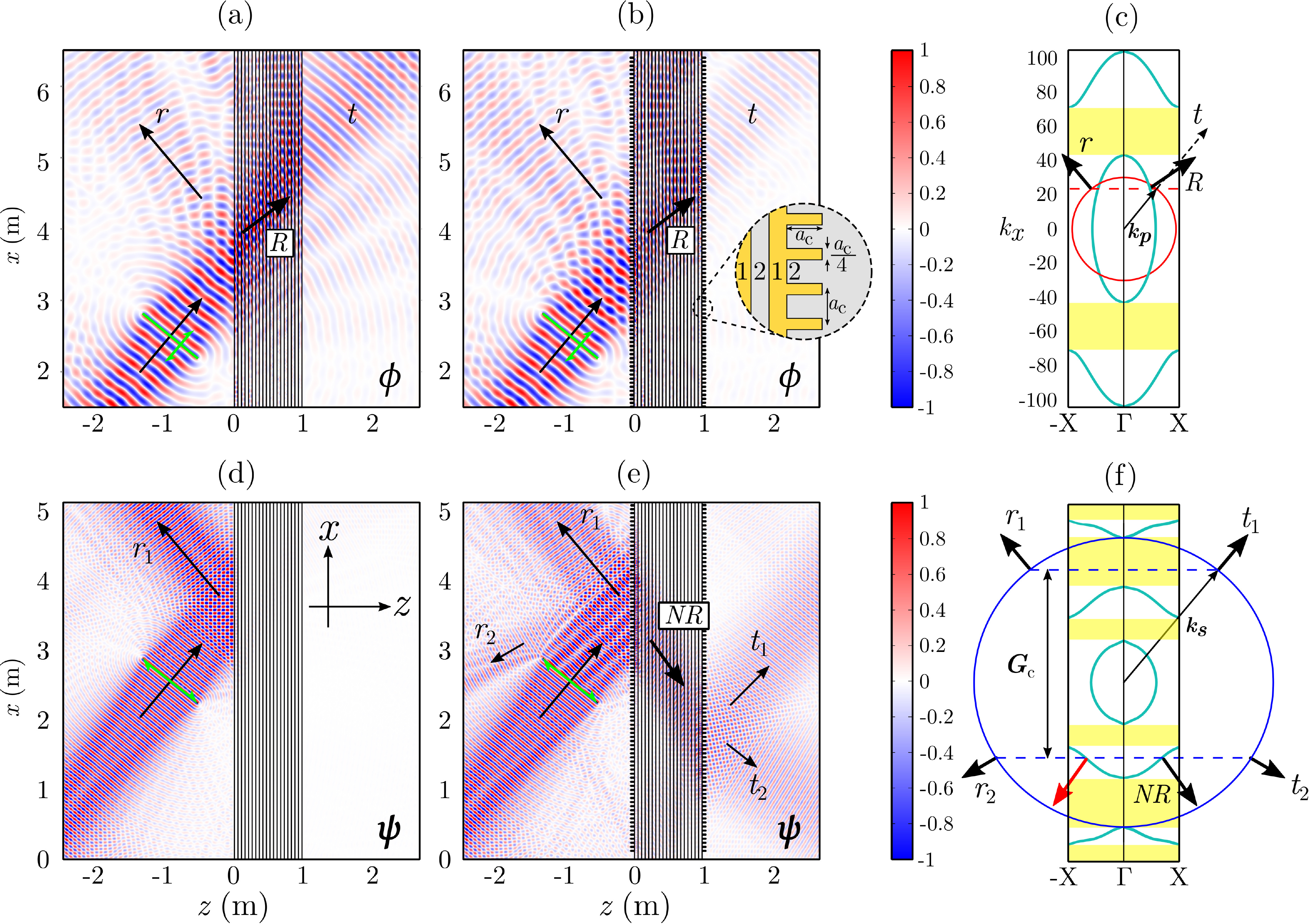}
    \caption{Splitting of SV- and P-waves through selective diffraction by surface corrugation: (a) Total compressional potential field, $\phi$. Compressional waves are preferentially excited by forcing the line source perpendicular to its direction, at a non-dimensional frequency $\Omega = 24$. The line forcing is oriented at $51^{\circ}$ to the laminate axis, and marked by a green line, with the particle motion shown by the half-headed arrow. Here no surface corrugation is present. The laminate consists of 10 bilayers such that $a_1=a_2=5\si{\centi\meter}.$ (b) Similar to (a) but with corrugation present on both sides of laminate in the form of extrusions of layer 1, with dimensions $a_c \times 0.25a_c$, where $a_c = 5.7\si{\centi\meter}$ is the periodicity of the corrugation, shown in inset. (c) Isofrequency contours of the laminate (green) with isocircle of incident P-wave in material 2 surrounding the laminate shown in red. The construction line (tangential component of $\boldsymbol{k}_{p}$ shown in dashed-red) intersects the contours and so conventional refraction occurs at an angle defined by the normal to the intersection. The reflected, refracted and transmission directions are shown by the arrows marked with $r$, $R$ and $t$ respectively, corresponding to those shown in (a-b). (d) Similar to (a) but for SV excitation, which are preferentially excited by applying a forcing parallel to the line direction.  At this frequency and angle there exists a bandgap (yellow regions); the construction line in (f) of $\boldsymbol{k}_{s}$ (dashed blue line) does not intersect any of the laminates contours and so only reflection occurs, shown by $r_1$. (e) Similar to (b) but for SV excitation. Negative refraction and higher order reflection occurs due to the intersection of the translated construction line via momentum transfer of $\boldsymbol{k}_{s}-\boldsymbol{G}_c$, as predicted by the isofrequency contours shown in (f), which has the same vertical scale as (c). Here the isocircle of the incident SV-wave in material 2 is shown by the blue circle, with the directions of reflection, negative refraction and, by reciprocity, transmission shown by the arrows labelled $r_1$, $r_2$, $NR$, $t_1$ and $t_2$ respectively. The red arrow highlighted in (f) shows a non-causal wavevector; the energy flow within the crystal is towards the source and so cannot be excited\cite{foteinopoulou2005electromagnetic}. Low reflecting boundary conditions are applied on the edges of the domain in (a,b,d,e).}
    \label{fig:Isofreq}
\end{figure*}

Inspired by the ability to achieve negative refraction through surface corrugation in optical systems\cite{Lu2007}, we extend this to elasticity showing that the resulting negative refraction is not only selective by angle, as in the optical case, but also by wave type, due to the propagation of the two elastic body waves at different speeds. The idea rests upon momentum transfer due to the periodic nature of the surface corrugation, on account of wavevectors in a periodic structure being defined up to a reciprocal lattice vector $\boldsymbol{G}_c$. This concept is commonly exploited in grating couplers \cite{chuang2012physics}, 2D photonic crystals\cite{foteinopoulou2005electromagnetic} and recently for manipulating surface elastic Rayleigh waves\cite{chaplain2020tailored}.

The design of the surface corrugation relies upon interpreting the isofrequency contours gained from the TMM method. Given an incident P(SV)-wave of wavevector of $\boldsymbol{k}_{p(s)}$ at an angle $\theta_{1p(s)}$ to the laminate axis (Figure~\ref{fig:Motivation}(b)), there is then a Bloch wave component ${k}_{z} = \cos\theta_{1p(s)}|\boldsymbol{k}_{p(s)}|$ in the $z$-direction of periodicity, along with a vertical component parallel to the laminate interface ${k}_{x} = \sin\theta_{1p(s)}|\boldsymbol{k}_{p(s)}|$. As such, for each frequency a contour is defined over the angles of incidence. Example isofrequency contours at the non-dimensional frequency $\Omega = 24$ are shown in Figure~\ref{fig:Isofreq}(c,f) for the cases of incident P- and SV-waves respectively. The $z$-direction is periodic such that $\Gamma \equiv \boldsymbol{k}_{z} = 0$ and $\pm X \equiv \boldsymbol{k}_{z} = \pm \frac{\pi}{(a_1+a_2)}$.

Figure~\ref{fig:Isofreq} shows the effect of selective diffraction by angle, frequency and wave-type for elastic materials with arbitrary properties such that $c_{2p}/c_{1p} \sim 4.3$ and $c_{2s}/c_{1s} \sim 2.8$, similar to the ratios achievable between some polymers\cite{mason1958physical}. We stress here that given the non-dimensional approach taken in constructing the TMM, the dispersion relations for laminates composed of arbitrary materials can quickly be characterised. This highlights an additional benefit of the adopted methodology, since comparisons between materials with unusual elastic properties (e.g. metamaterials with effective properties\cite{kadic2012practicability}) can be considered. The laminate is composed of 10 unit bilayer strips such that $a_1 = a_2 = 5\si{\centi\meter}$, embedded in a material with the same properties as material 2, as shown in Figure~\ref{fig:Motivation}(b). As such, additional to the contours shown in Figure~\ref{fig:Isofreq}(c,f) we show the isotropic contours of P and SV waves in material 2, shown in red and blue respectively. Determining the design of the corrugation for splitting SV- and P-waves then rests on interpreting the intersections of the tangential component of the incident wave at the interface (sometimes referred to as the construction line\cite{foteinopoulou2005electromagnetic}) with the contours of the periodic structure. 

In the presence of a vertical periodic corrugation, the vertical wavevector component, $k_x$, of the incident wave is then also defined up to a reciprocal lattice vector of the vertical grating such that $\boldsymbol{G}_c = [0,2\pi/a_c]$ with $a_c$ the vertical period (recall the $(z,x)$ coordinate system). As such, translations of the construction line, by integer multiples of $\boldsymbol{G}_c$, result. The desired wave splitting is then achieved by ensuring this induced momentum "kick" results in selective negative refraction for only one wave type. This is predicted from intersections of the translated construction lines with the isofrequency contours of each wave type, and the resulting directions of energy flow (normal to this intersection).

In Figures~\ref{fig:Isofreq}(a,b) we show the compressional potential for an incident wave, predominantly of compressional motion due to the perpendicular forcing of the line source, at $51^{\circ}$ to the laminate axis. Figure~\ref{fig:Isofreq}(a) shows the case where the laminate has no corrugation on the outer interfaces, whilst Figure~\ref{fig:Isofreq}(b) shows the effect of adding a surface corrugation of periodicity $a_c = 5.7\si{\centi\meter}$ on both sides of the laminate. The corrugation takes the form of a periodic array of extrusions of the first material, i.e. blocks of dimension $a_1 \times 0.25a_1$, shown by the inset in Figure~\ref{fig:Isofreq}(b). We note that the exact form of the grating, and therefore the guided modes it supports, is somewhat arbitrary. The selective diffraction mechanism only rests on the periodicity of the structure. As such the corrugation could easily be, for example, a gentle sinusoidal variation in thickness; the coupling efficiency will however be effected by the depth of the grating, just as in its optical counterparts\cite{marchetti2019coupling}. Interpreting the contours in Figure~\ref{fig:Isofreq}(c) shows that the similarities in (a,b) are expected; at this angle the construction line intersects the a contour of the laminate, with or without the presence of the grating. The angle of refraction is then determined by the normal to this intersection point. As such the P-waves remain unaffected by the surface corrugation.

In stark contrast to this, are Figures~\ref{fig:Isofreq}(d,e), which show the effect without and with the corrugation on the shear potential. Shear waves are predominantly excited by forcing the line source in a direction parallel to it. Figure~\ref{fig:Isofreq}(d) shows the shear potential at the same frequency and angle as before, with no corrugation present. There is no intersection with the construction line and the contours within the laminate, shown in Figure~\ref{fig:Isofreq}(f), and as such total reflection occurs at the interface. However when the corrugation is included, as in Figure~\ref{fig:Isofreq}(e), we see that incorporating momentum transfer by the grating by the \textit{subtraction} of a reciprocal lattice vector $G_{c} = [0,2\pi/a_c]$ (often attributed to the Umklapp mechanism\cite{foteinopoulou2005electromagnetic,chaplain2020ultrathin}) from the incident $\boldsymbol{k}_s$ results in an intersection of the \textit{translated} construction line with the laminate contours. This is shown in Figure~\ref{fig:Isofreq}(f). Determining the normal to this intersection then gives the angle of \textit{negative refraction} for the incident SV-wave, which is clearly visible in Figure~\ref{fig:Isofreq}(e). Additionally to this there is a higher order reflection at the interface, also shown, again due to momentum transfer via the corrugation. The angles of transmission on the other side are shown, arising due to reciprocity. Hence by incorporating the grating the transmitted SV- and P-potential fields are split. 


\section{Conclusions}
We have theoretically shown that it is possible to separate the shear-vertical and compressional elastic wave components through selective diffraction, by incorporating surface corrugation on the outer interfaces of a periodic bilayered elastic laminate. The design rests on obtaining its corresponding isofrequency contours, and then incorporating the desired momentum transfer from the vertical periodic stratification. To demonstrate this effect we used block extrusions of the first medium in the bilayer to form a grating on either side of the laminate. We note that the form of the grating is somewhat arbitrary, all that is required is there is some periodicity which provides the relevant momentum "kick" to the incident wave. Indeed the grating need not be symmetric on either side of the laminate, providing a further avenue for the manipulation of elastic waves. The simplicity of this then has advantages over say, tailored phased metasurfaces, as all that is required to predict the effects are the isofrequency contours of the layered structure, which we obtain through a non-dimensionalised version of the TMM. The demonstrated separation of the elastic potential fields was realised for two gratings which enabled selective negative diffraction for only P-waves (Fig.~\ref{fig:Hero}) and only for SV-waves (Fig.~\ref{fig:Isofreq}). Additionally, for certain frequency ranges and angles of incidence, both wave types will be affected and experience different angles of refraction, depending on the band structure of the distinct wave types. As such the presented surface corrugated laminate structures form a class of elastic grating couplers capable of physical effects further than their scalar wave counterparts.   
\begin{acknowledgments}
G.J.C gratefully acknowledges financial support from the EPSRC in the form of a Doctoral
Prize Fellowship. The support of the UK EPSRC through grant EP/T002654/1 is acknowledged as is that of the ERC H2020 FETOpen project BOHEME under grant agreement No. 863179.

\end{acknowledgments}

\section*{Data availability}
The data that support the findings of this study are available from the corresponding author upon reasonable request.

\appendix

\section{\label{sec:Append1} Parameters for Figure 1.}
The elastic parameters of the bilayers and surrounding elastic material defined in Figure~\ref{fig:Hero} are the same as those utilised in Figure~\ref{fig:Isofreq}. All that differs are the dimensions of the corrugation; the periodicity of which is defined as $a_c = 12\si{\centi\meter}$, with width $a_c/2$ and height $a_c/4$. The non-dimensional frequency of excitation in this case is $\Omega = 40$ at an angle of $30^{\circ}$ with respect to the laminate normal.

\section{\label{sec:AppendA} Transfer Matrix Method}
Here we detail the TMM by firstly nondimensionalising the governing equations, in order to circumvent numerical instabilities which arise in the TMM. Particularly, this is unstable at high frequencies and layer thicknesses (the large \textit{f-d} problem\cite{dunkin1965computation}). We reformulate the governing wave equations in terms of the shear wave speed in the first layer, $c_{1s}$. This allows us to express the TMM formalism entirely in terms of ratios of the speeds and material properties in each layer. We introduce the length scale $l$ such that $Z = zl$ and $X = lx$. Therefore the equation for the shear potential in the first layer reads
\begin{equation}
    \nabla_{Z}^2\boldsymbol{\psi}_{1} + \Omega_{1}^2\boldsymbol{\psi}_{1} = 0,
\end{equation}
where $\nabla_{Z}^2 = \frac{1}{l^2}\left(\partial_Z^2 + \partial_X^2\right)$ and $\Omega_{1} = \frac{l\omega}{c_{1s}}$. Letting subscripts denote the layer order, and implicitly incorporating the re-scaling, the rest of the potentials are then governed by 
\begin{align}
    \begin{split}
        \nabla^2\phi_{1} + (\Omega_{1}\xi_{1p})^2\phi_{1} = 0, \\
        \nabla^2\boldsymbol{\psi}_{2} + (\Omega_{1}\xi_{2s})^2\boldsymbol{\psi}_{2} = 0, \\
        \nabla^2\phi_{2} + (\Omega_{1}\xi_{2p})^2\phi_{2} = 0,
    \end{split}
\end{align}
where $\xi_{i\delta}^2 = \frac{c_{1s}^2}{c_{i\delta}^2}$, with $i = 1,2$ and $\delta = p, s$. Two separate versions of the TMM then arise, one for incident P waves, and one for incident SV waves. Solutions then arise in the form of a superposition of left and right propagating waves in each layer. Considering the case of a purely incident P wave, the angles can all be expressed in terms of the incident angle $\theta_{1p}$ from \eqref{eq:Snell}. As such the solutions can be written
\begin{align}
    \begin{split}
        \boldsymbol{\psi}_{1} &= (A_{1s}e^{i\xi_{1p}\Omega_{1}\alpha_{0}X}e^{i\Omega_1\beta_1Z} + A_{2s}e^{i\xi_{1p}\Omega_{1}\alpha_{0}X}e^{-i\Omega_1\beta_1Z})\uvec{j}\\
        \phi_{1} &= A_{1p}e^{i\xi_{1p}\Omega_1\alpha_0X}e^{i\xi_{1p}\Omega_1\beta_{0}Z} + A_{1p}e^{i\xi_{1p}\Omega_1\alpha_0X}e^{-i\xi_{1p}\Omega_1\beta_{0}Z} \\
        \boldsymbol{\psi}_{2} &= (B_{1s}e^{i\xi_{1p}\Omega_{1}\alpha_{0}X}e^{i\xi_{2s}\Omega_1\gamma_1Z} + B_{2s}e^{i\xi_{1p}\Omega_{1}\alpha_{0}X}e^{-i\xi_{2s}\Omega_1\gamma_1Z})\uvec{j} \\
        \phi_{2} &= B_{1p}e^{i\xi_{1p}\Omega_1\alpha_0X}e^{i\xi_{2p}\Omega_1\gamma_{2}Z} + B_{2p}e^{-i\xi_{1p}\Omega_1\alpha_0X}e^{i\xi_{2p}\Omega_1\gamma_{2}Z},
    \end{split}
\end{align}
where $\alpha_{0} = \sin\theta_{1p}$, $\beta_{0} = \cos\theta_{1p}$, $\beta_{1} = \cos(\sin^{-1}(\xi_{1p}\alpha_{0}))$, $\gamma_{1} = \cos(\sin^{-1}(\zeta_{2s}\alpha_{0}))$, $\gamma_{2} = \cos(\sin^{-1}(\zeta_{2p}\alpha_{0}))$ where $\zeta_{i\delta} = \frac{c_{i\delta}}{c_{1p}}$. The amplitudes $A_{1(2)}$ correspond to the right(left) propagating waves in the first layer, with $B_{1(2)}$ to those in the second. In order to maintain the non-dimensional formulation we then construct the vector $\boldsymbol{W}'_{i} = (\frac{w}{i\Omega_1},\frac{u}{i\Omega_1}, \frac{\sigma_{xz}}{\Omega_1^2c_{1p}^2\rho_{1}},\frac{\sigma_{xz}}{\Omega_1^2c_{1p}^2\rho_{1}})^{\text{T}}$.
Factorising the $e^{i\xi_{1p}\Omega_{1}\alpha_{0}x}$ terms allows us to write 
\begin{equation}
    \boldsymbol{W}'_{1} = C_{1}D_{1}\boldsymbol{A},
    \label{eq:TMM1}
\end{equation}
where $\boldsymbol{A} = (A_{1p},A_{2p},A_{1s},A_{2s})^{\text{T}}$, and $D_{1}$ is the propagator matrix for layer 1 such that $D_{1} = \text{diag}(\exp{\big[i\xi_{1p}\Omega_1\beta_0Z},{-i\xi_{1p}\Omega_1\beta_0Z},{i\Omega_1\beta_1Z},{-i\Omega_1\beta_1Z}\big])$. Similarly in the second layer
\begin{equation}
    \boldsymbol{W}'_{2} = C_{2}D_{2}\boldsymbol{B},
    \label{eq:TMM2}
\end{equation}
where $\boldsymbol{B} = (B_{1p},B_{2p},B_{1s},B_{2s})^{\text{T}}$, and $D_{2}$ is the propagator matrix for layer 2 such that $D_{2} = \text{diag}(\exp{\big[i\xi_{2p}\Omega_1\gamma_2Z},{-i\xi_{2p}\Omega_1\gamma_2Z},{i\xi_{2s}\Omega_1\gamma_1Z},{-i\xi_{2s}\Omega_1\gamma_1Z}\big])$, with the forms of $C_{1}$ and $C_{2}$ shown below. We focus on a unit strip consisting of two layers, centred at $Z = 0$ with the first layer of width $a_1$ and the second of $a_2$, such that $l = a_1+a_2$. Without loss of generality we make the coordinate shift $Z \rightarrow Z+a_1$, and as such at the edge of layer 1
\begin{align}
\begin{split}
    \boldsymbol{W}'_{1}(-a_1) = C_1D_1(-a_1)\boldsymbol{A} \\ 
    \implies \boldsymbol{A} = C_1^{-1}\boldsymbol{W}'_{1}(-a_1).
\end{split}
\end{align}
Similarly at the opposite edge of layer 2,
\begin{align}
\begin{split}
    \boldsymbol{W}'_{2}(a_2) = C_2D_2(a_2)\boldsymbol{B} \\ 
    \implies \boldsymbol{B} = [C_2D_2(a_2)]^{-1}\boldsymbol{W}'_{2}(a_2).
\end{split}
\end{align}
Evaluating \eqref{eq:TMM1} and \eqref{eq:TMM2} at the centre $Z = 0$ i.e. the interface between layers 1 and 2 gives
\begin{align}
    \begin{split}
        \boldsymbol{W}'_{1}(0) &= P_1 \boldsymbol{W}'_{1}(-a_1), \\
        \boldsymbol{W}'_{2}(0) &= P_2 \boldsymbol{W}'_{2}(a_2), 
    \end{split}
\end{align}
where $P_{1} = C_1D_1(0)C_1^{-1}$ and $P_{2} = C_2D_2(0)D_2^{-1}(a_2)C_2^{-1}$. Enforcing the continuity equations requires that 
\begin{align}
    \begin{split}
        \boldsymbol{W}'_{1}(0) &= \boldsymbol{W}'_{2}(0) \\
        \implies \boldsymbol{W}'_{2}(a_2) &= P_2^{-1}\mathds{1}P_1\boldsymbol{W}'_{2}(-a_1).
    \end{split}
\end{align}
Enforcing the Bloch condition, given the structure is periodic in $Z$, then gives
\begin{align}
    \begin{split}
        \boldsymbol{W}'_{2}(a_2) = e^{i\xi_{1p}\Omega_1\beta_{0}}\boldsymbol{W}'_{1}(-a_1).
    \end{split}
\end{align}
As such the dispersion relation arises from solving the eigenvalue problem
\begin{equation}
    (Q - e^{i\xi_{1p}\Omega_1\beta_{0}}\mathds{1})\boldsymbol{W}'_{1}(-a_1) = 0,
\end{equation}
which is satisfied when $\det[Q - e^{i\xi_{1p}\Omega_1\beta_{0}}\mathds{1}] = 0$, with $Q = P_2^{-1}\mathds{1}P_1$. A similar approach is adopted for an incident SV-wave, with each angle being written in terms of $\theta_{1s}$. We note that despite the advantages of nondimensionalisation, for large incident angles the matrices can become close to singular due to small entries within $P_2$. Inaccuracies surrounding this can be alleviated by using the pseudoinverse.
\subsection{$C_{i}$ Matrices}
The forms of $C_{i}$ are given below, in terms of the non-dimensional ratios of speeds, $\xi$ and $\zeta$ defined above and in terms of the non-dimensional density $\Upsilon = \frac{\rho_2}{\rho_1}$. Where the matrix components are too lengthy they are shown separately in the form of $C^{(pq)}_{i}$ with $p,q$ denoting the row and column, in the $i^{th} = 1,2$ corresponding elastic layer.
\begin{widetext}
\begin{align}
    C_{1} &= \begin{pmatrix}
    \xi_{1p}\beta_0     & -\xi_{1p}\beta_0      & -\xi_{1p}\alpha_{0}       & -\xi_{1p}\alpha_0 \\ \\
    \xi_{1p}\alpha_0    & \xi_{1p}\alpha_0      & \beta_1                   & -\beta_1          \\ \\
    C^{(31)}_1 & C^{(32)}_1 & 2\xi_{1p}^3\alpha_0\beta_1 & -2\xi_{1p}^3\alpha_0\beta_1 \\ \\
    -2\xi_{1p}^4\alpha_0\beta_0 & 2\xi_{1p}^4\alpha_0\beta_0 & \xi_{1p}^2\left(\left(\xi_{1p}\alpha_0\right)^2 - \beta_1^2\right) & \xi_{1p}^2\left(\left(\xi_{1p}\alpha_0\right)^2 - \beta_1^2\right)
    \end{pmatrix}, \\ \\ \\
    C_{2} &= \begin{pmatrix}
    \xi_{2p}\gamma_2 & -\xi_{2p}\gamma_2 & -\xi_{1p}\alpha_0 & -\xi_{1p}\alpha_0 \\ \\
    \xi_{1p}\alpha_0 & \xi_{1p}\alpha_0  & \xi_{2s}\gamma_{1} & -\xi_{2s}\gamma_1 \\ \\
    C^{(31)}_2 &  C^{(32)}_2 & 2\zeta_{2s}^2\Upsilon\xi_{1p}\xi_{2s}\alpha_0\gamma_1 & -2\zeta_{2s}^2\Upsilon\xi_{1p}\xi_{2s}\alpha_0\gamma_1 \\ \\
   -2\zeta_{2s}^2\Upsilon\xi_{1p}\xi_{2p}\alpha_0\gamma_2 & 2\zeta_{2s}^2\Upsilon\xi_{1p}\xi_{2p}\alpha_0\gamma_2 & \zeta_{2s}^2\Upsilon\left(\left(\xi_{1p}\alpha_0\right)^2-\left(\xi_{2s}\gamma_1\right)^2\right) & \zeta_{2s}^2\Upsilon\left(\left(\xi_{1p}\alpha_0\right)^2-\left(\xi_{2s}\gamma_1\right)^2\right)
    \end{pmatrix}, \\ \\ \\
    \begin{split}
        C^{(31)}_1 &= 2\xi_{1p}^4 + \frac{2i\xi_{1p}^3\beta_0}{\Omega_1} -\xi_{1p}^2, \\
        C^{(32)}_1 &= 2\xi_{1p}^4 - \frac{2i\xi_{1p}^3\beta_0}{\Omega_1} -\xi_{1p}^2, \\
        C^{(31)}_2 &= \Upsilon\xi_{1p}^2\left(\alpha_{0}^2(2\zeta_{2s}^2-\zeta_{2p}^2)-\gamma_2^2\right)+2\zeta_{2s}^2\Upsilon\left(\frac{i\xi_{2p}\gamma_2}{\Omega_1}+\xi_{2p}^2\gamma_2^2\right), \\
        C^{(32)}_2 &= \Upsilon\xi_{1p}^2\left(\alpha_{0}^2(2\zeta_{2s}^2-\zeta_{2p}^2)-\gamma_2^2\right)+2\zeta_{2s}^2\Upsilon\left(\xi_{2p}^2\gamma_2^2 -\frac{i\xi_{2p}\gamma_2}{\Omega_1}\right).
    \end{split}
\end{align}
\end{widetext}

\section{\label{sec:AppendB} Elastic Grating Coupler}

As elucidated, the term grating coupler is almost exclusively used to describe electromagnetic devices, particularly prevalent in integrated silicon photonic devices and fibre optics \cite{taillaert2002out,chuang2012physics,marchetti2017high,marchetti2019coupling}. In elasticity, coupling achieved by periodic gratings typically describes excitation of surface acoustic waves utilising piezoelectric coupling \cite{white1965direct,zhang1993characteristic,golan1991grating}. Certain elastic devices have utilised diffractive effects from non-uniform gratings to introduce frequency dependent time delays for shear waves\cite{coquin1965theory}. We show here, as an additional example, that conventional grating coupler modalities can be achieved in the elastic setting. The result is that, due to the presence of designed grating in a local region along the laminate, wave energy can enter and be trapped within the laminate due to subsequent total internal reflections.

\begin{figure}
    \centering
    \includegraphics[width = 0.475\textwidth]{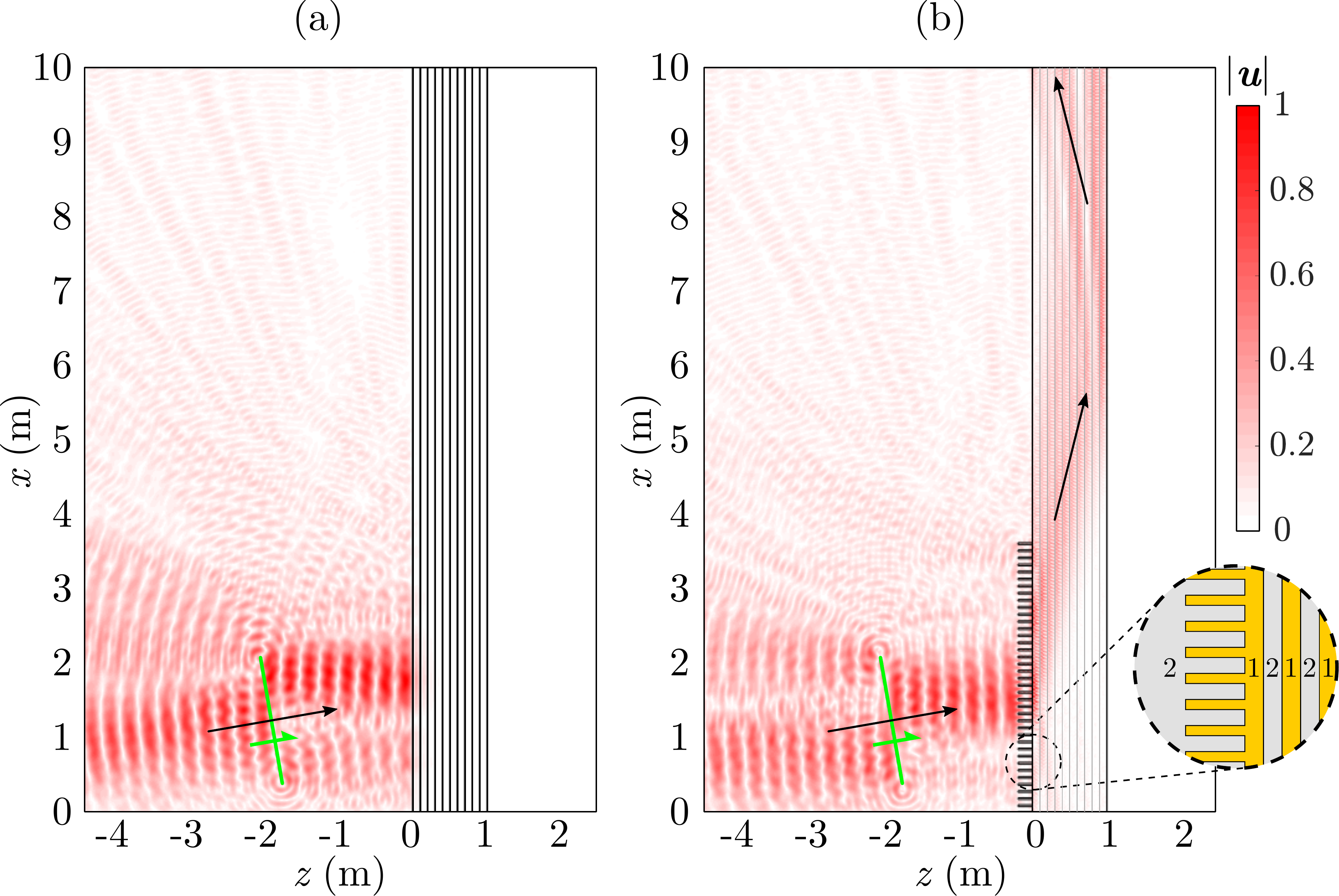}
    \caption{Elastic Grating Coupler: (a) Total solid displacement field ($|\boldsymbol{u}|$) excited by perpendicular line forcing (green line, similar to Fig.~\ref{fig:Isofreq}(a)) at $9.8^{\circ}$ to the laminate axis. The non-dimensional frequency of excitation is $\Omega = 50$. There is no surface grating present on the laminate structure and as such no elastic energy penetrates into the laminate, a schematic of which is shown in Figure~\ref{fig:Motivation}. (b) Total solid displacement field for the same scenario as in (a), except there is a grating present on the outer interface of the laminate (the internal structure is drawn in light grey as to not obscure the field). The presence of the grating incorporates a transfer of crystal momentum related to the periodicity of the grating (shown in inset). The wave is subsequently trapped in the laminate waveguide by total internal reflections. The laminates are composed of bilayer strips of elastic materials labelled 1,2 (coloured gold and silver respectively in the inset), similar to the schematic shown in Figure~\ref{fig:Motivation}.}
    \label{fig:TIR}
\end{figure}

Figure~\ref{fig:TIR} shows a direct comparison to optical devices, whereby the presence of a periodic surface structuring enables incident radiation to couple to a waveguide. The structure is similar to that considered in Figure~\ref{fig:Isofreq}, all that has altered is the grating; the periodicity of which is defined as $a_c = 9.5\si{\centi\meter}$, with width $a_c/2$ and height $a_c/4$. There are 40 cells present in the simulation of Figure~\ref{fig:TIR}(b), only present on the left interface of the structure. Figure~\ref{fig:TIR}(a) shows the total solid displacement field which results given a perpendicular forcing of the line source, oriented at $9.8^{\circ}$ to the laminate normal; for the chosen frequency of excitation there is no supported mode within the waveguide and as such reflection from the interface occurs. Introducing a periodic grating in the form of a surface corrugation on the outer interface of the laminate allows, via the transfer of crystal momentum, excitation of a diffractive order capable of coupling with a waveguide mode. This is shown in Figure~\ref{fig:TIR}(b). Furthermore the grating is only introduced in the region desired to couple incident radiation; no grating is present anywhere else along the array and as such, by total internal reflection the waveguided mode is now trapped within the laminate. 


%

\end{document}